\documentclass[twocolumn,prl,showpacs,preprintnumbers,amsmath,amssymb]{revtex4}
\usepackage{graphicx}% Include figure files
\usepackage{dcolumn}% Align table columns on decimal point
\usepackage{bm}% bold math
\usepackage{subfigure}

\begin{document}

\title{Skyrmionic textures in chiral magnets}

\author{Ulrich K. R\"o\ss ler, Andrei A. Leonov, Alexei N.\ Bogdanov}

\address{IFW Dresden, Postfach 270116, D-01171 Dresden, Germany}

%\author{Ulrich K. R\"o\ss ler, Andrei A. Leonov, Alexei N.\ Bogdanov}

%\address{IFW Dresden, Postfach 270116, D-01171 Dresden, Germany}

%\ead{u.roessler@ifw-dresden.de}

\begin{abstract}
{
In non-centrosymmetric magnets, the chiral Dzyaloshinskii-Moriya exchange stabilizes
Skyrmion-strings as excitations which may condense into multiply modulated phases.
Such extended Skyrmionic textures are determined by the stability of 
the localized 'solitonic' Skyrmion cores and their geometrical incompatibility 
which frustrates regular space-filling. We present numerically exact 
solutions for Skyrmion lattices and formulate basic properties of 
the Skyrmionic states.
}
\end{abstract}

%\address{IFW Dresden, Postfach 270116, D-01171 Dresden, Germany}

%\maketitle

\maketitle

% \clearpage
%

\vspace{5mm}

\section{Introduction}

\vspace{2mm}

During the last years the investigation 
of chiral modulations and Skyrmionic states,
both in bulk non-centrosymmetric magnetic materials 
and nanomagnetic systems (see e.g. \cite{PRL01,Nature06,Bode07,
Pappas09,Muhlbauer09,Lee09} and bibliography
in \cite{Nature06,Uniaxial09}), 
has become a hot topic.
Recently different specific modulated states
have been discovered  in non-centrosymmetric
magnets and multiferroics 
including the manifestation 
of Skyrmionic states in the
cubic helimagnet MnSi 
\cite{Pappas09,Muhlbauer09,Lee09}.
Such Skyrmionic states have been predicted
two decades ago \cite{JETP89}, and
subsequent theoretical studies within
the standard Dzyaloshinskii theory \cite{Dz64}
have revealed a number of novel modulated states 
in magnetic materials
with intrinsic and induced chirality
\cite{PRL01,Nature06,JMMM94,Filippov97}.
In this contribution, we outline
main results on chiral Skyrmionic
states in the low temperature region
\cite{JMMM94} and near the ordering
temperature \cite{Nature06} derived
within the standard phenomenological
theory \cite{Dz64}.
We also present new numerically exact 
solutions for Skyrmion lattices.
We formulate basic properties of
Skyrmionic states and elucidate 
physical mechanisms underlying 
their formation and stability.

\section{Chiral flux-lines: the building blocks 
of Skyrmionic matter}

\vspace{2mm}

According to Dzyaloshinskii \cite{Dz64},
the magnetic energy density of
a non-centrosymmetric ferromagnet
can be written as 
\begin{equation}
w=A \sum_{i,j}(\partial_i M_j)^2
-\mathbf{M}\cdot\mathbf{H}+w_D(\mathbf{M})+w_0(\mathbf{M}),
\label{density}
\end{equation}
and includes the exchange 
stiffness with constant $A$,
the Zeeman energy, and
chiral Dzyaloshinskii-Moriya (DM)
coupling ($w_D$).
$w_0(\mathbf{M})$ combines internal energy
contributions independent on gradients of the 
magnetization.

\begin{figure*}%[h]
%\vspace{2mm}
%\begin{minipage}{38pc}
\includegraphics[width=17cm]{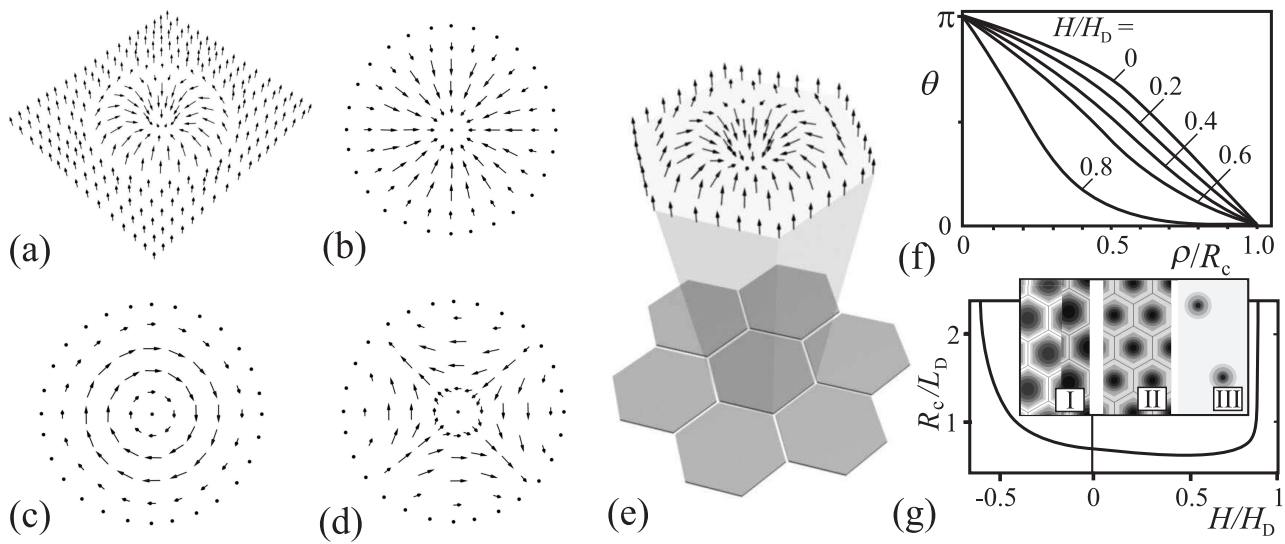}
\caption{
\label{f1} 
Cross-section through an
isolated Skyrmion in 
ferromagnets with $C_{nv}$ symmetry (a).
Projections of the magnetization onto
the basal plane in the Skyrmion core 
for ferromagnets with
$C_{nv}$ (b), $D_n$ (c), and
$D_{2d}$ (d) symmetries.
A hexagonal Skyrmion lattice for
ferromagnets with $C_{nv}$ symmetry
(e).
Evolution of the magnetization 
$\theta (\rho/R_c)$
with increasing magnetic field
(f) (adopted from Ref. \cite{JMMM94}).
Equilibrium size $R_c$ of the Skyrmion
cell as a function of an applied
magnetic field ($L_D = 4\pi A/D$) (g). 
Inset shows
the distribution of the magnetization
in the Skyrmion lattice for
$H/H_D$ = 0 (I), 0.6 (II), 0.8 (III)
 ($H_D = M D^2/(2A)$).
}
%\end{minipage}
%\hspace{2pc}%
%\vspace{2mm}
\end{figure*}

Chiral modulations arise as a result of
the competition between the exchange
and DM interactions \cite{Dz64},
and a sufficiently strong
magnetic field compresses them
into localized two-dimensional 
vortex-like structures (Fig. \ref{f1} (a))
\cite{JETP89,JMMM94}.
These chiral axisymmetric structures have 
a non-trivial topology and smooth non-singular 
cores with definite sizes.
The DM couplings are described in continuum models 
of magnetic materials by so-called Lifshitz invariants,
energy contributions linear in first spatial derivatives
of the magnetization, 
$\Lambda_{ij}^{(k)} = M_i\partial_k M_j-M_j\partial_k M_i$
\cite{Dz64}. For MnSi and other B20 compounds
belonging to the crystallographic class T, 
$w_D=D\,(\Lambda_{yx}^{(z)}+\Lambda_{xz}^{(y)}+\Lambda_{zy}^{(x)})$
=$D\,\mathbf{M}\cdot \mathrm{rot}\mathbf{M}$ 
(where $D$ is a Dzyaloshinskii constant).
For uniaxial non-centrosymmetric classes the
functionals $w_D$  are given in Ref. \cite{JETP89}.

For the sake of simplicity we include
into model (\ref{density}) only
these basic interactions \textit{essential}
to stabilize  Skyrmionic states, and
neglect other less important energy 
contributions (as magnetic anisotropy,
stray-fields, magneto-elastic coupling).

For low temperatures $M=\mathrm{const}$, and
the equations minimizing energy (\ref{density})
with $w_0=0$
include solutions for axisymmetric localized
structures $\psi = \psi(\phi)$,
$\theta = \theta(\rho)$,
where
$\mathbf{M}=M(\sin\theta\cos\psi;
\sin\theta\sin\psi;\cos\theta))$
and the spatial variable
$\textit{\textbf{r}}=
(\rho\cos\varphi;\rho\sin\varphi;z)$
is written in cylindrical coordinates
(Fig. \ref{f1}).
The solutions $\psi (\phi)$
are determined by crystal
classes of the system \cite{JETP89}.
Particularly,
$\psi  = \phi$ for C$_{nv}$ symmetry,
$\psi = \phi - \pi/2$ for
$D_n$ and cubic T classes, and
$\psi = \pi/2 - \phi$ for
$D_{2d}$ symmetry (Fig. \ref{f1},
(a-d)).
The polar angle $\theta(\rho)$
is derived from equation

\begin{align}
A \left( \frac{d^2 \theta}{d \rho ^2} 
+ \frac{1}{\rho}\frac{d \theta}{d \rho}\right.
 -&\left.\frac{1}{\rho ^2}\sin \theta \cos \theta  \right)\nonumber\\
&-\frac{D}{\rho} \sin^2 \theta
- \frac{H}{2M}\sin \theta =0,
\label{eq}
\end{align}
with the boundary conditions 
$\theta(0)=\pi,\, \theta(\infty)=0$.
This yields the solutions for isolated Skyrmions
which are radially stable at well-defined sizes $L \propto |D|/H$ \cite{JETP89}
(Fig. \ref{f1} (f,g)).
The chiral interactions play
the crucial role to stabilize Skyrmions.
In centrosymmetric systems ($D = 0$) 
such solutions are radially
unstable and collapse spontaneously under
the influence of the applied magnetic field
or intrinsic short-range interactions
\cite{JETP89,JMMM94}.

Commonly in nonlinear field models 
Skyrmions arise due to specific
invariants described
by higher order spatial derivatives
(so-called \textit{Skyrme mechanism}).
In condensed-matter systems there
are no physical interactions providing
such energy contributions.
Hence, chiral couplings
present the unique mechanism to stabilize 
Skyrmionic textures in ordered condensed matter 
systems described by a large class 
of nonlinear field models \cite{Nature06}.
This singles out 
\textit{chiral} condensed-matter
systems with Lifshitz-type of invariants (including non-centrosymmetric
magnets, multiferroics, and chiral liquid
crystals) into a particular class of
materials with Skyrmionic states.
The Skyrmions in chiral magnets can be thought of 
as isolated filaments within spatially homogeneous phases.
Isolated Skyrmions remind 
Abrikosov vortices in
type II superconductors or 
thread-like textures in nematic
liquid crystals \cite{Nature06}.
Contrary  to these defected patterns
with singularity in the core, the 
distribution of the order parameter in Skyrmions is smooth 
(Fig.~\ref{f1}).

\begin{figure*}%[h]
%\vspace{2mm}
%\begin{minipage}{38pc}
\includegraphics[width=17cm]{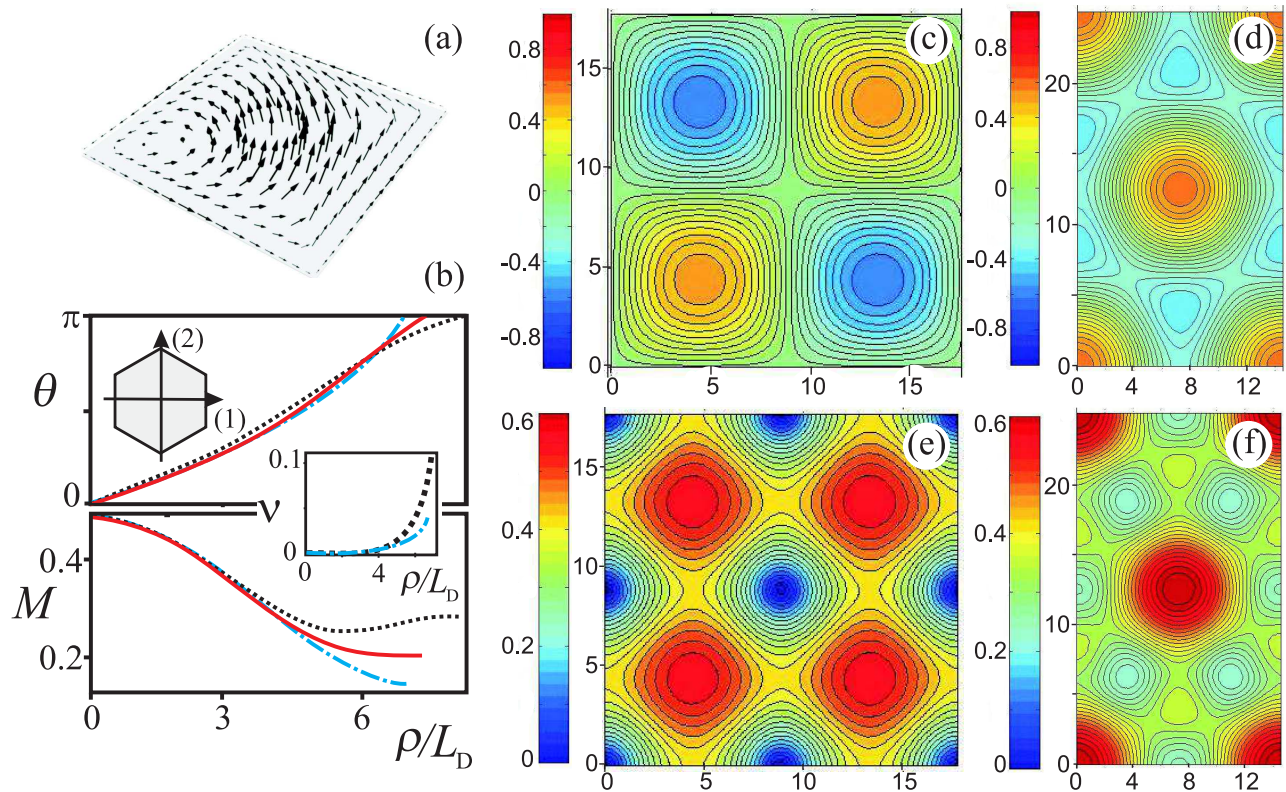}
\caption{\label{f2} 
(Color online) 
A "half-Skyrmion" lattice cell
(a).
Profiles $\theta_i (\rho)$ 
and $M (\rho)$ (b) in a
hexagonal cell for two
different directions $i$ through the core
(dashed-dotted line ($i=1$) and
 dotted line ($i=2$))
are plotted
together with the corresponding
profiles $\theta_0 (\rho)$ and $M_0 (\rho)$
for the circular cell approximation (solid line).
Inset in (b) shows
$\nu_i = |\theta_i - \theta_0|/\theta_0$
as functions of $\rho$.
Contour plots of $M_z(x,y)$ (c,d)
and $M(x,y)$ (e,f) for a square
(c,e) and hexagonal (d,f) lattices.
The results are derived
for reduced parameters
of model (\ref{density}) near the ordering
temperature:
$\tilde{a} = a(T-T_C)A/D^2$ = 0.23,
$\tilde{b} = bA/D^2$ = 0.05.
}
%
%\end{minipage}
%\hspace{2pc}%
\end{figure*}

\section{Skyrmion lattices versus helicoids}
\vspace{2mm}

For strong DM interactions isolated Skyrmions condense
into lattices (Figs. \ref{f1} (e),
\ref{f2} (c-f)). These 2D modulated textures
are alternatives to common
one-dimensional (\textit{helical}) modulations.
In the model with a fixed magnetization
modulus appropriate for low temperature 
the equilibrium parameters of
the Skyrmion lattice have been
calculated in circular cell
approximation \cite{JMMM94}.
At zero field the equilibrium period
of the lattice is close to the helicoid period
($2R_c \approx L_D$). Near the critical
field $H_c = 0.8132$  the lattice
transforms into a system of isolated
Skyrmions by infinite growth of the period
and localization of the core
(Fig.~\ref{f1} (f), (g)) \cite{JMMM94}.

Near the ordering temperature,
the magnetization modulus 
becomes small ($M \ll 1$)
and strongly depends on the 
applied field and temperature.
A brute-force minimization
of the functional (\ref{density})
with $w_0(\mathbf{M})=a(T-T_c)M^2+bM^4$
in zero field 
yields solutions for 
a "half-Skyrmion" staggered
structure and hexagonal lattice
with radial variation
of the magnetization modulus in the lattice cells ( Fig. \ref{f2}).
\cite{Nature06}.
The Skyrmion lattices are characterized by a strong
variation of the cell sizes and transformation
of their structures near cell boundaries.
However, they preserve axisymmetric 
distribution of the
magnetization near the cell center.
This remarkable property is due to
specific energetics of the Skyrmions.
"Double-twist" rotation of the magnetization
near the Skyrmion core leads to larger energy
reduction than in "single-twisted" helical phases
while edge areas of the cell have larger
energy density than the helical states
\cite{Nature06}.
This explains the unusual axial symmetry
of the cell cores and their stability.
The condensation of Skyrmions creates 
spatially inhomogeneous twisted phases 
as a result of space-filling by these 
multi-dimensional solitonic objects.

\section{Summary}

\vspace{2mm}
% The competition between chiral
% and other magnetic interactions
% can yield a large variety of 
%
Skyrmionic textures reveal
common features imposed by
the fundamental physical mechanisms
underlying their stability.
Summarizing the findings of
previous papers \cite{Nature06,JETP89,JMMM94}
and results of this contribution
we  formulate basic properties
of extended states built from chiral Skyrmion 
solutions.
(i) In magnetic materials lacking
inversion symmetry chiral DM
interactions stabilize isolated
\textit{Skyrmions}, axisymmetric
localized structures with a fixed
rotation sense and definite shape and
sizes.
(ii) Strong DM coupling leads
to the condensation of Skyrmions
into lattices (Figs. \ref{f1},  \ref{f2}).
The effective interactions between
the Skyrmions are weak relative to 
energies determining their radial stability. 
Hence, in real materials, the detailed arrangement 
of Skyrmions into lattices and their stability strongly 
depend on additional magnetic couplings 
such as anisotropies, applied and dipolar 
stray fields, temperature, and static disorder.
%
% Dense packing of the 
% cylindrical Skyrmion strings in a uniaxial setting will appear 
% as hexagonal arrangement, as observed in MnSi \cite{Muhlbauer09}.
% But, under influence of thermal fluctuations
% Skyrmionic states may also appear as disordered 
% liquid or glass-like states \cite{Pappas09}. 
%
%
(iii) Skyrmions, both isolated and bounded,
are characterized by axisymmetrical
distribution of the magnetization
in their cores (Fig. \ref{f2} (b-f)).

The main features of chiral Skyrmions highlighted
in this paper provide a basis for detailed 
analysis of recent experimental findings
\cite{Pappas09,Muhlbauer09,Lee09} and
 allow to establish relations
between the classical phenomenological model (1) and 
other theoretical approaches
(see \cite{Nature06,Muhlbauer09}
and bibliography of these papers).

\end{document}